\documentclass{amia}
\usepackage{lipsum}
\usepackage{subfigure}
\usepackage{wrapfig}
\usepackage{xcolor}
\usepackage{algorithm}
\usepackage{multirow}
\usepackage{makecell}
\usepackage{algpseudocode}
\begin{document}

\title{Facilitating Federated Genomic Data Analysis by Identifying Record Correlations while Ensuring Privacy}

\author{Leonard Dervishi$^1$, Xinyue Wang$^2$, Wentao Li$^3$, Anisa Halimi$^4$, Jaideep Vaidya$^2$, Xiaoqian Jiang$^3$, Erman Ayday$^1$}

\institutes{
    $^1$ Case Western Reserve University, Cleveland, OH; $^2$Rutgers University, Newark, NJ; $^3$UTHealth, Houston, TX; $^4$IBM Research Europe, Dublin, Ireland
}

\maketitle

\section*{Abstract}
\vspace{-5pt}
\textit{With the reduction of sequencing costs and the pervasiveness of computing devices, genomic data collection is continually growing. However, data collection is highly fragmented and the data is still siloed across different repositories. Analyzing all of this data would be transformative for genomics research. However, the data is sensitive, and therefore cannot be easily centralized. Furthermore, there may be correlations in the data, which if not detected, can impact the analysis. 
In this paper, we take the first step towards identifying correlated records across multiple data repositories in a privacy-preserving manner. 
The proposed framework, based on random shuffling, synthetic record generation, and local differential privacy, allows a trade-off of accuracy and computational efficiency. An extensive evaluation on real genomic data from the OpenSNP dataset shows that the proposed solution is efficient and effective.
}

\vspace{-5pt}
\section*{Introduction}\label{sec:introduction}
\vspace{-10pt}
In genomic research, collaboration between multiple researchers often produces more accurate outcomes and powerful statistics. 
On the other hand, the use of genomic data in collaborative studies has serious privacy implications, as it includes information about an individual's phenotype, ethnicity, family memberships, and disease conditions, which might be highly sensitive to the study participants. 
Existing techniques to overcome such privacy concerns in collaborative studies include (i) meta-analysis, which allows collaborators to only exchange statistics with each other to obtain global (more reliable) statistics; (ii) cryptographic solutions,
which allow collaborators to conduct statistical analyses over (encrypted) federated datasets; (iii) and differential privacy (DP)-based solutions, in which collaborators can exchange statistics or research datasets between each other under certain privacy guarantees.

One overlooked step in privacy-preserving collaborative studies is the identification of records (samples) that should be included and/or filtered for the collaborative study. Such a step is crucial to make sure that the collaborative research results are obtained using high quality data (e.g., after certain bias is removed or after focused groups are identified). This step may include (i) identifying related samples and filtering them (as a part of the quality control process) in genome-wide association studies (GWAS); (ii) identifying related samples (family members) across different datasets to conduct studies that only involve family members, such as family-based GWAS; and (iii) identifying similar patients (e.g., who carry a set of rare variants) to study a particular disease or mutation. 
Existing privacy-preserving solutions for genomic data processing (e.g., cryptographic solutions or DP-based solutions) are not feasible to achieve this because (i) cryptographic solutions are not efficient for large scale analyses (identification of samples to be included or removed from a collaborative study involves analyzing the pairwise relationships between all the records in the combined dataset), and (ii) DP-based solutions are based on obfuscating the shared data, and such obfuscation typically leads to high inaccuracies in identifying the records to be included or removed from the collaborative study. 

As a first step to tackle this crucial challenge, we propose efficient privacy-preserving techniques to identify correlated samples across distributed genomic datasets. 
Among several potential record correlations, in this work, we mainly focus on identifying kinship relationships since the existence of close family members in GWAS greatly impacts the collaborative study outcome. 
Concurrently, sample relatedness is one of the main steps in quality control (QC), which is widely used to clean the genomic data before performing any type of genomic study such as GWAS. 
Locally identifying and eliminating related samples at each researcher's dataset is not sufficient because two or more datasets that include no kinship relationships in isolation, might include such relationships in the combined dataset due to cross correlations across them. On the other hand, analyzing the combined dataset to identify such relationships should not violate the privacy of research participants. 

We consider a client-server environment, in which researchers, who want to conduct the collaborative study, share metadata with a server so that the server can analyze and identify the kinship relationships across the samples in their datasets, and hence identify which records to keep and which records to filter out in their collaborative study. 
We have two main goals: (i) accurate identification of kinship coefficients across datasets and (ii) restricting the privacy risk to below a baseline. 
To achieve (ii), we formulate the privacy risk (considering membership inference) and show that the proposed scheme can keep this risk below a baseline that occurs due to sharing summary statistics (which is accepted by many institutions, including the NIH~\cite{nih}).

To simultaneously compute the kinship coefficients across samples from different datasets efficiently and  provide privacy, we propose a technique, in which researchers (collaborators) lightly synchronize to decide on a common set of single nucleotide polymorphisms (SNPs) to be shared with the server and randomly shuffle the SNPs to be shared so that the server cannot link the SNPs in the received datasets to their actual IDs (i.e., un-shuffle the SNPs in the received datasets). To further improve the privacy (i.e., reduce the un-shuffling probability of the server), we also explore addition of synthetic samples to researchers' datasets and using a variant of local differential privacy (LDP) before sharing the partial datasets with the server. 
Therefore, for the given privacy goal (e.g., keeping the privacy risk below a baseline), the proposed scheme allows a trade-off between accuracy (for correct identification of the kinship relationships) and computational efficiency. 
To analyze and quantify the privacy of the proposed scheme, we model an efficient technique to un-shuffle the received datasets, showing the risk of un-shuffling at the server, and quantify the risk for membership inference (by using the data that is obtained as a result of un-shuffling) via a power analysis. 
Our results on a real-life genomic dataset show that the proposed scheme simultaneously achieves (i) high accuracy for the identification of the kinship relationships (over 95\% accuracy when more than 250 SNPs are provided), (ii) low un-shuffling risk (less than 40\% un-shuffling accuracy when more than 40\% additional synthetic samples are added and LDP variant is applied to the partial noisy dataset), and (iii) low membership inference risk (a membership inference power less than 0.5 when the un-shuffling accuracy is 40\% or less).
\vspace{-5pt}
\section*{Related Work}\label{sec:relatedwork}
\vspace{-10pt}
Several privacy-preserving techniques have been developed to support 
genomic testing and genome-wide association analysis.
Frikken et al.~\cite{frikken2009practical} proposed methods for DNA string matching, Bruekers et al.~\cite{bruekers2008privacy} and Baldi et al.~\cite{baldi2011countering} developed methods for paternity tests. Ayday et al.~\cite{ayday2013protecting} and Namazi et al.~\cite{namazi2019dynamic} provided methods for privacy-preserving genetic susceptibility tests. Considering the nature of genome data, methods built on cryptographic techniques, such as oblivious transfer and homomorphic encryption, are not scalable to large-scale genomic data. Several different taxonomies of privacy-preserving GWAS methods have been proposed. Blatt et al.~\cite{blatt2020secure} developed a secure outsourced GWAS schema based on homomorphic encryption. Cho et al.~\cite{cho2018secure} proposed a secure GWAS protocol using multiparty computation requiring no collusion among the parties. 
In ~\cite{tkachenko2018large,johnson2013privacy}, differential privacy is utilized to alleviate the computation bottleneck. However, these methods come with a significant reduction in utility. 

Researchers have increasingly paid attention to privacy-preserving similar patients search, which is also relevant to our work.
In~\cite{jha2008towards} Jah et al. proposed three secure protocols built on oblivious transfer for privacy-preserving edit distance computation of genomic data. 
To alleviate the heavy computation cost of oblivious transfer, Wang et al.~\cite{wang2015efficient} developed an algorithm to approximate the edit distance for the human genome by transforming the computation to a private set difference size problem. However, the public reference genomic sequence required by their method can affect the accuracy. 
Asharov et al.~\cite{asharov2018privacy} developed an approximation function for edit distance by pre-processing the whole genome sequences into fragments. Their protocol by using garbled circuits can return the exact k-closest records in 98\% of the queries.
There is some work on outsourcing similar patient search. 
In~\cite{zhu2021privacy}, the data owner encrypts the genome data and the generated index with a hierarchical index structure. Once received the outsourced data, the cloud service provider can process the query efficiently using the proposed index merging mechanism based on bloom filter. 
Recently, Zhu et al.~\cite{zhu2021efficient} proposed a secure similar patient search mechanism based on a gBK-tree, edit distance approximation algorithm, and symmetric-key encryption. 

All aforementioned approaches to identify similar patients aim to identify similar genomic sequences to a given target sequence. Assuming the size of the dataset to conduct the search is $N$, these methods typically require comparisons between $N$ samples, and they still have the aforementioned limitations. 
On the other hand, in our problem setting, there is no target genome sequence; our proposed scheme aims to compare every pair of records across research datasets (and hence it requires a comparison between $N^2$ samples). Considering the scalability issues of the existing similar patient search protocols, they cannot be utilized to solve the problem we consider (which has higher computational requirements than the original similar patient search problem).
Our proposed solution offers an alternative efficient, privacy-preserving, and scalable solution for identification of record correlations across genomic datasets.

\vspace{-5pt}
\section*{Background Information}\label{sec:background}
\vspace{-10pt}
In this section, we briefly introduce the metrics used for kinship inference and local differential privacy.

\textbf{\textit{Kinship inference.}} 
To check for relatedness between individuals in genomic datasets, one can use any existing kinship metrics, such as KING coefficient~\cite{manichaikul2010robust}, Identity By Descent (IBS) allele estimation which is used in PLINK~\cite{purcell2007plink}, KIND~\cite{zhu2008unified}, and Graphical Representation of Relationship errors (GRR)~\cite{abecasis2001grr}. In this work, we use the KING kinship coefficient~\cite{manichaikul2010robust} due to its simplicity and efficacy even when the number of SNPs per user is low. Nevertheless, any other kinship metric can be applied to our proposed scheme.
Formally, the KING kinship coefficient between two individuals $i$ and $j$ is:
\vspace{-3mm}
\begin{equation}
\vspace{-1mm}
  \phi_{ij} = \frac{2n_{11}-4(n_{02}+n_{20})-n_{*1}+n_{1*}}{4n_{1*}},
 \label{eg:KINGcoefficient}
 \end{equation}
where $n_{11}$ is the total number of SNPs in which both individuals $i$ and $j$ are heterozygous, $n_{02}$ is the number of SNPs in which individual $i$ is homozygous dominant and individual $j$ is homozygous recessive, $n_{20}$ is the opposite of $n_{02}$ where individual $i$ is homozygous recessive and individual $j$ is homozygous dominant, and $n_{1*}$ and $n_{*1}$ are the count of SNPs in which individuals $i$ and $j$ are heterozygous, respectively. Based on KING, a kinship coefficient greater than the threshold $t^0 = 0.35$ implies duplicates or twins, less than $0.35$ but greater than $t^1 = 0.175$ implies parent-offspring relationship or full sibling (first degree relationships), less than $0.175$ but greater than $t^2 = 0.08$ implies second degree relatives, and so on~\cite{manichaikul2010robust}. In this work, we assume that SNPs are biallelic, and we set their values as 0, 1, and 2 representing the number of minor alleles they contain. 
  
\textbf{\textit{Local differential privacy (LDP).}} 
Differential privacy~(DP)~\cite{privacy:differentialprivacy} aims to preserve a record's privacy while publishing statistical information about a database. 
DP provides formal guarantees that the distribution of query results changes only slightly with the addition or removal of a single record in the database. 
DP guarantees that an algorithm behaves approximately the same on two neighboring databases $D1$ and $D2$ (that differ by a single record) as $Pr(\mathcal{K}(D1) \in S) \leq e^\epsilon \times Pr (\mathcal{K}(D2) \in S)$, where $\mathcal{K}$ is a randomized algorithm and $S$ is the output of the randomized algorithm ($\mathcal{K}$).

One variant of DP, called local differential privacy (LDP)~\cite{duchi2013local, kairouz2014extremal}, has been proposed to formalize privacy during individual data sharing. LDP is satisfied when an untrusted data collector cannot determine the original value of a data point from the reported (perturbed) value. 
Formally, for any two inputs $x_1$ and $x_2$ in the input space, and output $y$, an algorithm $\mathcal{K}$ satisfies $\epsilon$-LDP if
\vspace{-3mm}
\begin{equation}
\vspace{-1mm}
Pr[\mathcal{K}(x_1) = y] \leq exp(\epsilon) \times Pr [\mathcal{K}(x_2) = y].
\label{eg:localdifferential}
\end{equation}
$\epsilon$-LDP can be achieved by the randomized response mechanism~\cite{erlingsson2014rappor, wang2016using, wang2017locally}. 
In this mechanism, each individual shares his/her original value or an incorrect value based on probabilities determined by $\epsilon$. 
If the input set contains $d$ values, each individual shares his/her value correctly with probability $p = \frac{e^\epsilon}{e^\epsilon + d - 1}$ and shares each of the incorrect values with probability $q = \frac{1}{e^\epsilon + d - 1}$ to satisfy $\epsilon$-LDP~\cite{wang2017locally}. 

Direct application of the randomized response mechanism to achieve LDP significantly distorts the accuracy of the kinship coefficients, since the randomized response mechanism independently flips the values of SNPs at each sample. Therefore, in this work, we propose a variant of LDP (as will be discussed next) to preserve the kinship coefficients between individuals across different research datasets while preserving their privacy. 
\vspace{-5pt}
\section*{System and Threat Models}\label{sec:systemmodel}
\vspace{-10pt}

In this section, we present the general system model along with the threat model.

\textbf{\textit{System model.}} We consider a system that includes two parties (i) two or more researchers (collaborators) and (ii) a server as shown in Figure~\ref{fig:Framework} (details of the figure are discussed in the next section). The researchers are willing to identify related samples across their federated dataset in a privacy-preserving way, and all the computations are outsourced to the server.  
In the rest of this paper, we consider two researchers, but our methodology can be easily extended to multiple researchers. 
To check for sample relatedness across datasets, each researcher provides some metadata (generated from their local dataset, as will be discussed in detail in the next section) to the server. Then, using the received metadata, 
the server identifies the related samples across all the datasets and sends the sample IDs of the related individuals (samples) to each researcher. 
Based on the research study that the researchers are conducting, they can decide whether to use the related or unrelated samples.

\textbf{\textit{Threat model.}}
We assume honest researchers with legitimate research datasets. 
We also assume an honest-but-curious server that does the computations correctly, but on the other hand, it may try to infer sensitive information about the participants in the research datasets using the metadata it receives. 
There exist known privacy attacks, such as membership inference~\cite{homer2008resolving,wang2009learning}, attribute inference~\cite{humbert2013addressing,deznabi2017inference}, and de-anonymization attacks~\cite{gymrek2013identifying,humbert2015anonymizing} that exploit research results and/or partially provided datasets. 
In a membership inference attack, which is the most relevant attack for our considered scenario, the attacker (the server in our case) aims to determine whether a target individual is part of the dataset or not. 
In this work, the researchers' goal is to ensure that the server does not learn any additional information about the dataset participants besides what they can learn from the aggregate statistics about the dataset (release of such summary statistics is accepted by many institutions, including the NIH~\cite{nih}). 
In other words, researchers wants to ensure that the privacy risk due to the shared  metadata is lower than the risk due to sharing summary statistics. 
\vspace{-5pt}
\section*{Proposed Framework}\label{sec:proposedframework}
\vspace{-10pt}
Before getting into the details of the proposed framework, we provide the frequently used notations in Table~\ref{tab:symbols}.
\begin{table}[H]
\vspace{-10pt}
\centering
\begin{tabular}{cl|cl}
$R^i$ & Researcher $i$ & $U$ & Common random seed\\
$S$ & Server & $Q$ & Permutation vector generated from $U$\\
$n$ & Number of individuals in $D^i$ & $I$ & Set of SNP IDs shared with server $S$\\
$m$ & Number of SNPs shared with the server ($m=|I|$) & $n'$ & Number of synthetic samples added to $D^i$ \\
$D^i$ & The original dataset of $R^i$ & $M^i$ & Metadata sent by $R^i$ to the server $S$ \\
$\epsilon$ & privacy parameter & $x^j_k$ & State/value of SNP $k$ for individual $j$ \\
$r$ & Degree of relationship between two individuals (e.g., first)
& $t^r$ & KING coefficient threshold for $r$-th degree
\end{tabular}
\vspace{-10pt}
\caption{Frequently used symbols.}
\label{tab:symbols}
\vspace{-10pt}
\end{table}

\begin{wrapfigure}{r}{0.5\textwidth}
\centering
\vspace{-17pt}
\includegraphics[width=0.4\textwidth]{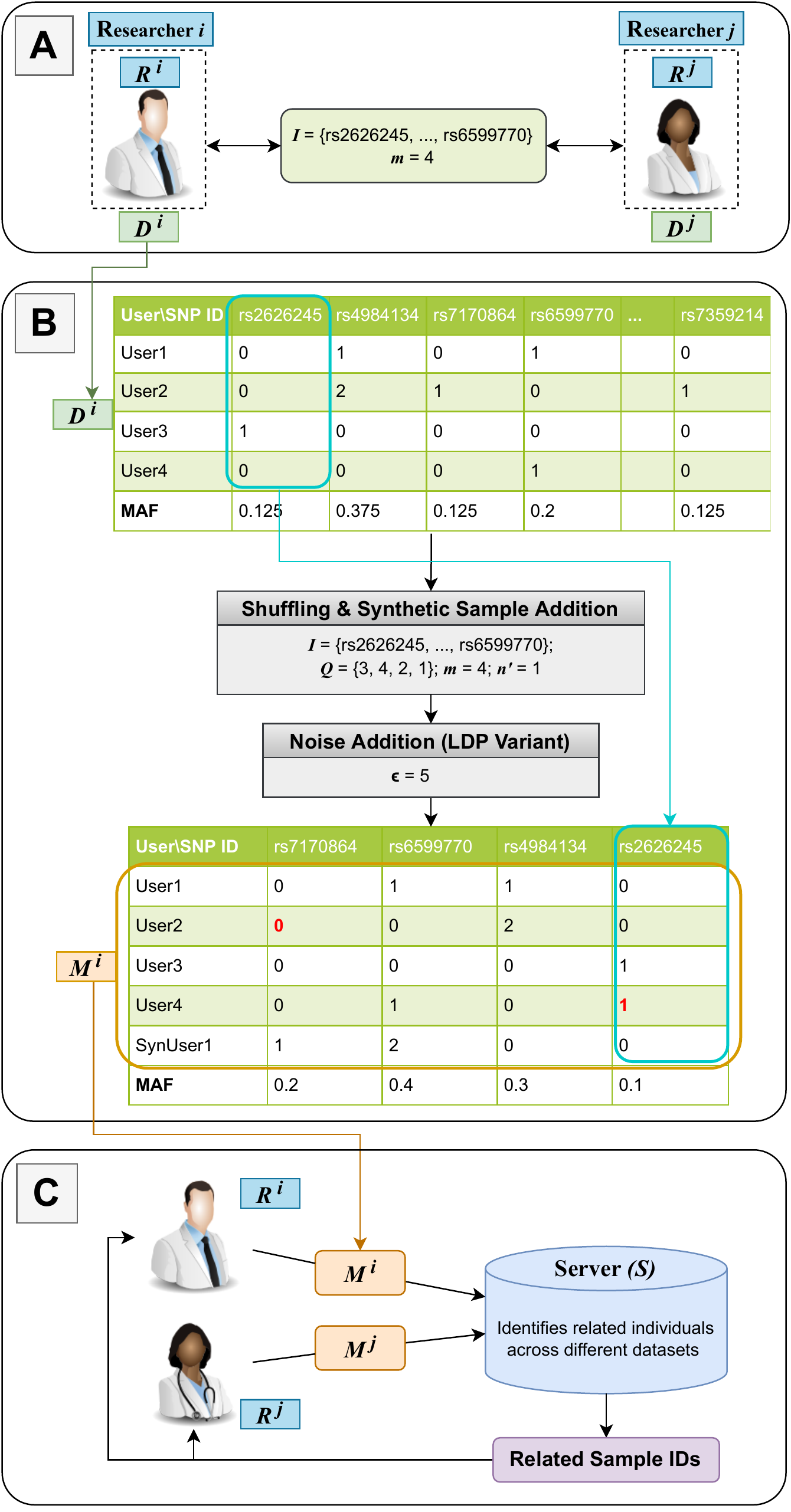}\vspace{-6pt}
\caption{\footnotesize{An overview of the proposed scheme. 
In (A), researchers coordinate to decide on the set of SNPs ($I$) that will be shared with the server and a common seed which is used to generate the permutation vector $Q$. (B) shows the generation of the metadata $M^i$ from the original dataset $D^i$ belonging to the researcher $R^i$. Initially, the researcher adds $n'$ synthetic samples and later shuffles the SNPs (columns) based on the permutation
in vector $Q$ (e.g., the values for the first SNP are moved to the last column as shown by the cyan colored arrow).
(B) also shows 
that some SNP values/states change for some users due to addition of noise to achieve the proposed LDP variant. The values in red denote the flipped SNP values.
In (C), each researcher $R^i$ sends their prepared metadata $M^i$ to the server, which computes the pairwise kinship coefficients among all users and sends back to the researchers the 
related IDs. 
}}
\label{fig:Framework}
\vspace{-27pt}
\end{wrapfigure}

To identify the kinship relationships between samples across different datasets, the proposed approach requires the researchers to share a part of their datasets (i.e., metadata) with the server in a privacy-preserving way. 
Here, each researcher has bi-fold goals: (i) metadata provided to the server should be useful to compute accurately the kinship coefficients between samples across different datasets; and (ii) metadata should not increase the baseline privacy risk that occurs due to sharing of aggregate statistics about the dataset.
Let $S$ denote the server, $D^i$ represent the original dataset of researcher $R^i$, and $M^i$ the metadata that is sent to the server from each researcher. 
In the following, we describe different steps of the proposed scheme (as also shown in Figure~\ref{fig:Framework}).

\textbf{\textit{Synchronization between the researchers.}} 
The proposed scheme initially requires the researchers to coordinate and decide on (i) the number of SNPs ($m$) and the set of SNP IDs $I=\{SNP_1, SNP_2, \ldots, SNP_m\}$ they will share with the server and (ii) a common seed $U$, as illustrated in Figure~\ref{fig:Framework}(A). Here, (i) is to make sure that the kinship coefficients between samples across datasets will be computed using the same set of SNPs, and hence they will be accurate. $m$ is a system parameter to control the relationship between the accuracy of the computed kinship coefficient, computational load at the server, and the privacy of dataset participants. On the other hand, (ii) is used by the researchers to shuffle their shared datasets in the same way, so that the server cannot identify the actual IDs of the shared SNPs.
Note that the set of SNP IDs $I$ should be selected in an optimal way such that the SNPs in $I$ 
have close MAF values to each other. This selection is to minimize the risk of un-shuffling at the server (we analyze this risk via evaluations). 

\textbf{\textit{Shuffling and synthetic sample addition at each researcher.}} 
Initially, each researcher generates a permutation vector $Q$ which is derived from the common seed $U$. Using the permutation vector $Q$, researchers shuffle the SNPs to be shared (i.e., shuffle the columns of the
partial dataset) as shown in Figure~\ref{fig:Framework}(B), so that the server receives these SNPs in a mixed order (i.e., the server will not know the IDs of the shared SNPs). 
On the other hand, since the shuffling is done using a common seed $U$ (and hence a common permutation vector $Q$) across the researchers, accuracy of kinship coefficients across the shuffled genomes remains intact. Therefore, the server, once it receives the shuffled datasets (metadata $M^i$) from the researchers, can accurately compute the kinship coefficient between all the pairs and identify the close relatives. We will conduct a privacy analysis for this shuffling idea by considering (i) the un-shuffling probability at the server and (ii) the risk of membership inference as a result of un-shuffling in the next section. 

As will be discussed, the risk of un-shuffling at the server increases as the server knows more about the statistics belonging to the original research dataset ($D^i$) of a researcher.\footnote{Exact statistics about a research dataset (e.g., MAF values or correlations between the SNPs) can be shared publicly and sharing of such information is acceptable by most institutions, including the NIH~\cite{nih}.} 
To improve the privacy of dataset participants (i.e., to decrease the probability of the server successfully un-shuffling the received datasets from the researchers), each researcher $R^i$ also adds $n'$ synthetic samples to their original dataset $D^i$ before sending the metadata ($M^i$) to the server. 
These synthetic samples are used to obfuscate the actual statistics in the original research dataset (e.g., MAF values) and to distort the correlation between SNPs. 
Researchers know which samples are synthetic, and hence addition of these synthetic samples does not decrease the accuracy of the process (i.e., if any kinship relationships between the synthetic records are identified, it is eventually ignored by the researchers). 
Adding more synthetic samples reduces the risk of un-shuffling at the server (as we will show experimentally), but high number of synthetic samples also increases the bandwidth requirement and the computation load at the server. 

\textbf{\textit{LDP variant on the shuffled dataset by preserving the kinship relationships.}} 
Adding synthetic samples decreases the un-shuffling (and hence, membership inference) risk, but as we will show, a high number of synthetic samples is required to significantly avoid the un-shuffling risk. As discussed, a high number of synthetic samples increases the bandwidth requirement, and it leads to additional (redundant) computations at the server. To optimize the computation load at the server, privacy risk, and accuracy of the computed kinship coefficients, we propose a hybrid approach that involves shuffling, addition of synthetic samples, and application of local differential privacy (LDP) for the shared SNPs with the server.

As discussed, LDP-based techniques rely on flipping the values of the shared SNPs based on a probability $q = \frac{1}{e^\epsilon + d - 1}$, where $d$ is the number of possible values/states that a SNP can take ($d$ = 3 in our case) and $\epsilon$ is the privacy parameter. 
Hence, privacy protection that comes with LDP results in a significant reduction of the accuracy of the kinship coefficients computed at the server. 
To minimize the accuracy loss in the computation of kinship coefficients between samples across different datasets, we propose a variant of LDP. 

In the standard $\epsilon$-LDP, there exist a probability $q$ for flipping a $0$ state (value) of a SNP to $2$ and vice versa.
Such alterations would significantly degrade the utility (accuracy of the kinship coefficient computation). 
Let $x^j_k$ represent the state for SNP $k$ of an individual $j$. 
In the proposed LDP variant, each individual $j$ shares their SNP value $x^j_k$ correctly with a probability $p = \frac{e^\epsilon}{e^\epsilon + d - 1}$ regardless of its state. On the other hand, 
\vspace{-0.2in}
\begin{itemize}
\setlength{\itemsep}{-0.01in}
    \item if $x^j_k = 0$ or $x^j_k = 2$, the state of SNP $k$ is flipped to $1$ with probability $q = \frac{2}{e^\epsilon + d - 1}$
    \item if $x^j_k = 1$, the state of SNP $k$ is flipped to $0$ or $2$ with probability $q = \frac{1}{e^\epsilon + d - 1}$
\end{itemize}
\vspace{-0.1in}
By doing so, we avoid a $0$ SNP value to be flipped to $2$ and vice versa. 
Although this variant of LDP provides weaker privacy guarantees compared to original LDP, it helps to preserve the kinship relationships between the samples (although noise is added to the SNPs of each sample completely independently). Besides, as we will also show via experiments, when combined with addition of synthetic samples, the proposed LDP variant provides high robustness against membership inference attacks. 
Finally, each researcher $R^i$ sends the metadata $M^i$, which consists of a partial noisy dataset, to the server.

\textbf{\textit{Kinship calculation at the server and identification of close relatives.}} 
As illustrated in Figure~\ref{fig:Framework}(C), the server computes the pairwise kinship coefficients between all users across different datasets by using the received metadata ($M^i$) from each researcher. Next, the server identifies all related individuals based on KING (kinship) coefficients and thresholds $t^r$ (that helps to classify a given kinship coefficient to a particular degree of kinship, i.e., $r$)~\cite{manichaikul2010robust}, where $r$ is the degree of relatedness (smaller degrees correspond to stronger relationships between samples).
Note that as the degree of relatedness increases (i.e., as $r$ increases), the accuracy of correctly identifying the nature of kinship between individuals decreases (which is a general drawback of kinship estimation using genomic data). 
Thus, for our evaluations, we consider up to second degree relationships ($r = 2$) to evaluate the accuracy of the proposed scheme. Finally, the server sends the list of all related individual IDs to the researchers.
\vspace{-5pt}
\section*{Privacy Analysis}\label{sec:privacy_analysis}
\vspace{-10pt}
The privacy risk in the proposed approach is equivalent to the probability of the server un-shuffling the shared SNPs in the metadata ($M^i$) and inferring their actual IDs. This is because if the server can successfully infer the IDs of the shared SNPs, then the dataset participants may be exposed to membership inference attacks (we also analyze the membership inference risk after the un-shuffling at the server). 
In the  following, we model the privacy risk due to un-shuffling as dataset reconstruction and analyze it thoroughly. 

To analyze the privacy risk, we assume that the server knows (i) the set of shared SNP IDs $I'$ (to consider the worst case, we assume $I'=I$, but in practice, $I'$ may be larger than $I$ if the server has some uncertainty about the set of shared SNPs), (ii) the MAF values of the SNPs in $I$ in the reference population, and (iii) pairwise correlations between the SNPs in $I$ based on public knowledge (i.e., a correlation table for each SNP pair representing the pairwise correlations between different states/values of the SNPs). 
The server, using its aforementioned background information and the metadata it receives from the researcher $R^i$, can generate a one-to-one match between the SNP IDs and the shuffled SNPs in the shared partial dataset $M^i$ (i.e., un-shuffle the shared SNPs) by comparing the MAF values and also using the additional correlation information between the SNPs. 

The optimal way to do this matching is using the Hungarian algorithm~\cite{kuhn1955hungarian} or a graph matching algorithm. For this, we can form a graph $G^T=\{V^T, E^T\}$, where the nodes (vertices) represent the SNPs of the metadata $M^i$ and the edges represent the correlations between the SNPs (the distances of the correlation tables). Each node also contains the MAF value of the SNP it represents. In the same way, we can form a graph $G^A=\{V^A, E^A\}$ for the SNPs in $I$ using the publicly available information. Then, we can compute the similarities between every possible pair of SNPs in both graphs, considering the MAF values and their connections. Then, by applying the Hungarian algorithm (or a graph matching algorithm), we can obtain a one-to-one match between the SNPs in $M^i$ and the ones in $I$. However, assuming $m$ SNPs in $M^i$ (and $I$), the running time of these methods is $O(m^3)$. Given their high complexity, in this work, we use a greedy algorithm. 
In the following, we discuss how the server may un-shuffle the SNPs in $M^i$ (metadata received from researcher $R^i$). The risk can be computed in the same way for other research datasets as well. 

To determine the real IDs of the shared SNPs in $M^i$, the server first computes the MAF values of each SNP in $M^i$.
Then, the server computes the pairwise correlations between the SNPs in $M^i$, i.e., it generates the correlation table for each SNP pair in $M^i$. 
The server matches the SNPs in $M^i$ with their IDs in $I$ in an iterative way, following the below steps:
\vspace{-12pt}
\begin{enumerate}
\setlength{\itemsep}{-0.015in}
    \item Compute the difference of the MAF values between all SNPs in $M^i$ and all SNPs in $I$. 
    \item Find the pair (e.g., ($M^i_j,I_l$) pair) with the minimum MAF difference (the one with the closest MAF values). 
    \item Assign the corresponding ID to the unknown SNP in $M^i$ (e.g., assign $I_l$ to the ID of SNP $j$ in $M^i$). 
    \item Pick randomly a SNP in $I$ (e.g., $I_b$) and compute the distance of the correlation table of the pair (e.g., ($I_l$, $I_b$)) with the correlation tables between the previously identified SNP ID (e.g., $M^i_j$) and all the other SNPs in $M^i$. 
    \item Find the pairs in $M^i$ that have the minimum correlation distance to the pair ($I_l$, $I_b$) (e.g., ($M^i_j$, $M^i_a$)). 
    \begin{enumerate}
    \setlength{\itemsep}{-0.01in}
    \vspace{-0.1in}
        \item If multiple pairs have close correlation distances to the pair ($I_l$, $I_b$), then find SNP with the closest MAF value to $I_b$ in the reference dataset (e.g., $M^i_a$). 
    \end{enumerate}
    \vspace{-0.03in}
    \item Assign the IDs of the corresponding SNPs (e.g., assign $I_b$ to the ID of SNP $a$ in $M^i$). 
    \item Remove the SNPs whose IDs are assigned from $I$ and $M^i$ (e.g., remove SNPs $M^i_j$, $M^i_a$ from dataset $M^i$, and SNPs $I_l$, $I_b$ from $I$). 
    \item Go back to step (4) until there are no SNPs left in $M^i$. 
\end{enumerate}
\vspace{-7pt}

As a result, the server un-shuffles the received metadata, inferring the actual IDs of the SNPs. 
The un-shuffled dataset $M^{i'}$ is prone to membership inference attacks (as discussed in the threat model). 
To model and analyze the membership inference risk, we assume that the server has access to a victim's SNP profile, which can be obtained from a blood or saliva sample (this is a common assumption to quantify the membership inference risk). 
To quantify the membership inference risk due to the un-shuffled dataset, we use a power analysis based on the hamming distance~\cite{halimi2021privacy}. 

\textbf{\textit{Power analysis based on hamming distance.}} To determine the match (closeness) between the genome of the target victim $i$ to any of the individuals' genomes in $M^{i'}$, we use the hamming distance (the minimum number of positions at which the genome sequences are different). We follow the same approach as in our previous work~\cite{halimi2021privacy} to quantify the power analysis using the hamming distance. 
First, we use $|A|$ individuals from a set $A$ that are not in the dataset $M^{i'}$. 
For each individual in $A$, we compute the hamming distance between the target $i$ and all individuals in the un-shuffled dataset $M^{i'}$ and identify the minimum hamming distance. 
Then, we identify the ``hamming distance threshold'' $\gamma$ as the $5\%$ false positive rate. 
Next, we use $|B|$ individuals from a set $B$ that are in the dataset $M^{i'}$. 
For each individual in $B$, we compute the hamming distance between the target $i$ and all individuals in the un-shuffled dataset $M^{i'}$ and identify the minimum hamming distance. 
Finally, we check what fraction of these $|B|$ individuals have minimum hamming distance that is lower than the threshold $\gamma$ (i.e., correctly identified as in $M^{i'}$).  
The privacy risk and the accuracy of this approach mainly depend on the number of shared SNPs, the amount of noise added to the metadata to achieve the proposed LDP variant (i.e., the privacy parameter, $\epsilon$), and the un-shuffling accuracy of the server. Thus, as shown in the next section, the researchers can fine-tune the privacy parameter of LDP and the number of shared SNPs in the research dataset to simultaneously achieve privacy against membership inference attacks, high accuracy for the computation of kinship coefficients, and low computational load at the server.

We compare the privacy risk of the proposed algorithm with the membership inference risk due to the sharing of the aggregate statistics about the researcher's dataset (e.g., MAF values). 
As discussed, sharing aggregate statistics is acceptable by many institutions, such as the NIH~\cite{nih}, and hence we refer to this risk as the ``baseline risk''. We use the likelihood-ratio test (LRT)~\cite{sankararaman2009genomic} to quantify it as follows: 

\textbf{\textit{Likelihood-ratio test.}} We assume that under the null hypothesis, the target $i$ is not part of the researcher's dataset $D^i$, and under the alternate hypothesis, the target $i$ is part of $D^i$. The overall log-likelihood ratio is computed as follows:
$LRT = \sum_{j=1}^l x_{i,j} log\frac{a_j}{pop_j} + (1-x_{i,j}) log\frac{1-a_j}{1-pop_j},$
where $x_{i,j}$ is the SNP $j$ of the individual $i$, $a_j$ is the MAF of SNP $j$ in $D^i$, and $pop_j$ is the MAF of SNP $j$ in the reference population. 

\vspace{-5pt}
\section*{Evaluation}\label{sec:results}
\vspace{-10pt}

To evaluate the proposed scheme, we used real genomic data from OpenSNP dataset~\cite{opensnp}, which consists of 28,000 SNPs from 942 samples. Additionally, following Mendel's law, we synthetically generated genome sequences for the first and second degree relatives of the users in the OpenSNP dataset. 

\begin{wraptable}{r}{0.45\textwidth}
\centering
\vspace{-14pt}
    \begin{tabular}{|c|c|c|c|c|}
    \cline{1-2} \cline{4-5}
    m & $Accuracy$ & & $\epsilon$ & $Recall$  \\ \cline{1-2} \cline{4-5}
    $50$ & $79\%$ & & $\epsilon=3$ & $86\%$ \\ \cline{1-2} \cline{4-5}
    $250$ & $95\%$ & & $\epsilon=4$ & $94\%$ \\ \cline{1-2} \cline{4-5}
    $500$ & $98\%$ & & $\epsilon=5$ & $98\%$ \\ \cline{1-2} \cline{4-5}
    $1000$ & $99\%$ & & $\epsilon=\infty$ & $98\%$ \\ \cline{1-2} \cline{4-5}
    $2500$ & $99\%$ & \multicolumn{3}{c}{} \\ \cline{1-2}
    \end{tabular}
    \caption{Kinship accuracy (for $\epsilon=5$) for varying number of SNPs ($m$) and recall (for $m=500$) for varying $\epsilon$ values.}
	\label{table:kinship_accuracy}
	\vspace{-10pt}
\end{wraptable}
We use the accuracy metric to measure the correctness of kinship identification using the KING coefficient~\cite{manichaikul2010robust}. 
We define the ``kinship accuracy'' as the fraction of the correctly identified kinship relationships over the total number of pairwise relationships.
As discussed, here, we only focused on the correct identification of the first and the second degree relatives. 
We use ``un-shuffling accuracy'' as the metric to measure the server's success in un-shuffling, which is computed as the fraction of the correctly matched SNP IDs over the total number of shared SNPs ($m$) in $M^i$. 
Furthermore, for the membership inference risk, we use power analysis based on hamming distance to measure the success of the server in identifying whether a target is in the research dataset $D^i$.

\begin{wrapfigure}{r}{0.4\textwidth}
\centering
\vspace{-3pt}
\includegraphics[width=0.4\textwidth]{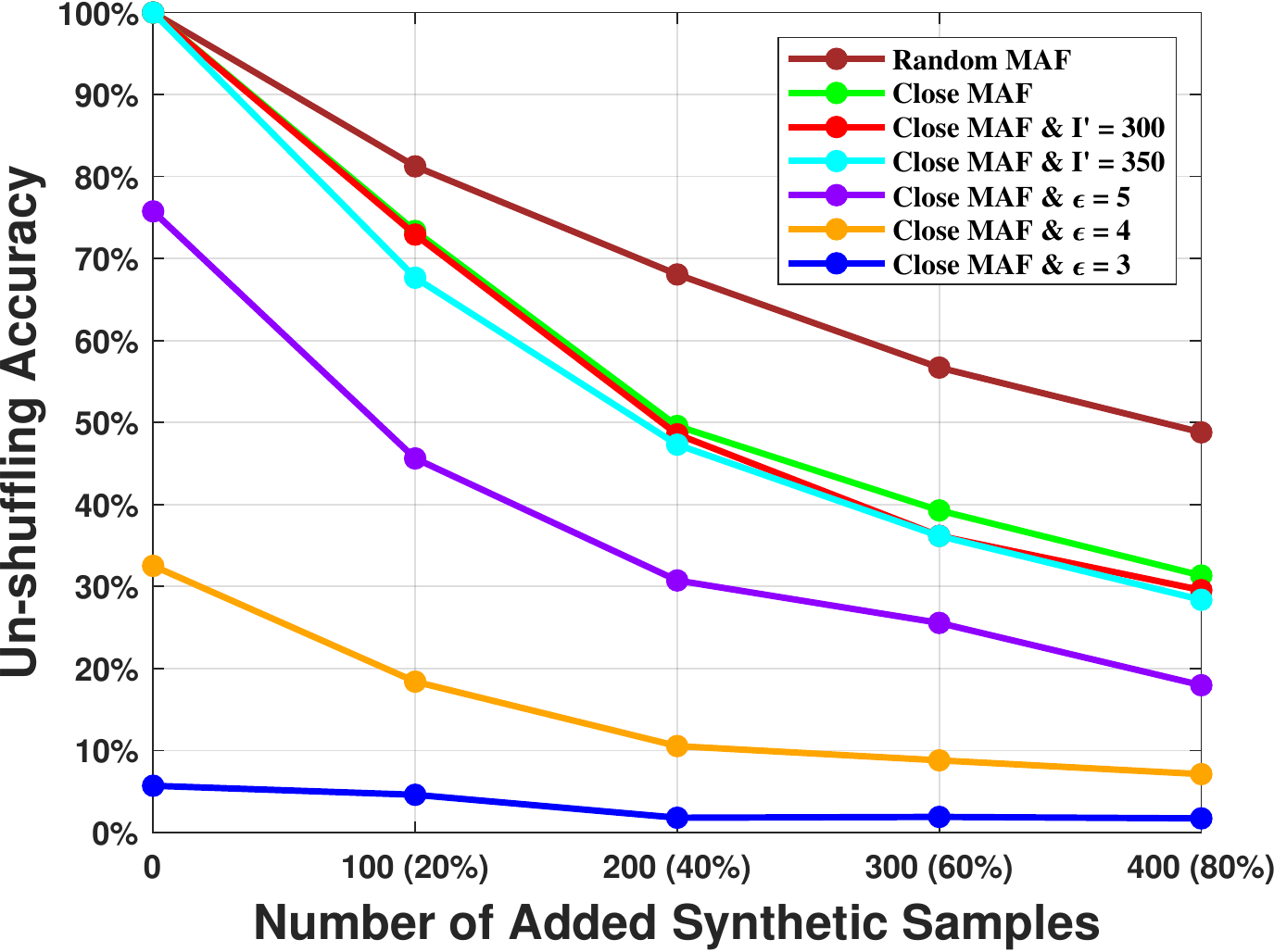}\vspace{-6pt}
\caption{\footnotesize{The server's accuracy in un-shuffling the shared SNPs in the metadata $M^i$ (i.e., identification of the real SNP IDs in $M^i$) for different scenarios. 
The value in parenthesis shows the fraction of added synthetic samples with respect to the number of original samples.
}}
\label{fig:Unshuffling_accuracy}
\vspace{-10pt}
\end{wrapfigure}

First, we explore the change in ``kinship accuracy'' with different number of SNPs (to compute the kinship coefficient) and different privacy parameter $\epsilon$ (to achieve the proposed LDP variant). 
For this, we generated a dataset which contains 100 unrelated individuals, along with 20 first degree and 10 second degree relatives of these individuals (the dataset contains 130 individuals in total). Next, we added noise to dataset entries for varying $\epsilon$ values ($\epsilon=\{3, 4, 5\}$), and finally, we computed the kinship accuracy for different number of SNPs in $M^i$ ($m=\{50, 250, 500, 1000, 2500\}$). The kinship accuracy for varying number of SNPs ($m$) when $\epsilon=5$ is shown in Table~\ref{table:kinship_accuracy}. In the same table, to show the effect of noise to achieve LDP, we also show the recall for varying $\epsilon$ values when $m=500$. 
For $\epsilon=5$, we observed an accuracy more than $95\%$ when the number of provided SNPs ($m$) is more than 250. We also observed that regardless of the value of $\epsilon$ the kinship accuracy is more than 99\% when the number of SNPs ($m$) to compute the kinship coefficient is more than $500$. 
The kinship accuracy decreases with decreasing $m$, as expected.  
We observed a similar trend for precision when $m$ decreases. In addition, for $m=500$, we obtained recall values of $86\%$, $94\%$, and $98\%$ for $\epsilon=3, 4$, and $5$, respectively, which shows that addition of more noise slightly degrades the utility of the shared data. For all above cases, we observed a slight decrease in accuracy if the provided SNPs have close MAFs.

Next, we investigate the accuracy of the server in un-shuffling (i.e., identifying the real IDs of the SNPs in the shared metadata $M^i$). 
We observed that attack performance (un-shuffling accuracy) decreases as $m$ increases. Thus, for this experiment, we set the number of provided SNPs in $M^i$ as $m=250$.
We also set the number of individuals in the dataset as $n=500$.
Since there is randomness in the addition of noise to achieve the LDP variant, in the generation of the seed, and addition of the synthetic samples, we conducted each experiment for 10 times and reported the average of the results. 
Note that (as we also show in Figure~\ref{fig:Unshuffling_accuracy}) the server successfully un-shuffles the entire set of SNPs in $M^i$ when researchers do not add any synthetic samples and 
noise to the shared metadata. 
This shows the requirement for the proposed countermeasures. 

\begin{wrapfigure}{r}{0.4\textwidth}
\centering
\vspace{-24pt}
\includegraphics[width=0.4\textwidth]{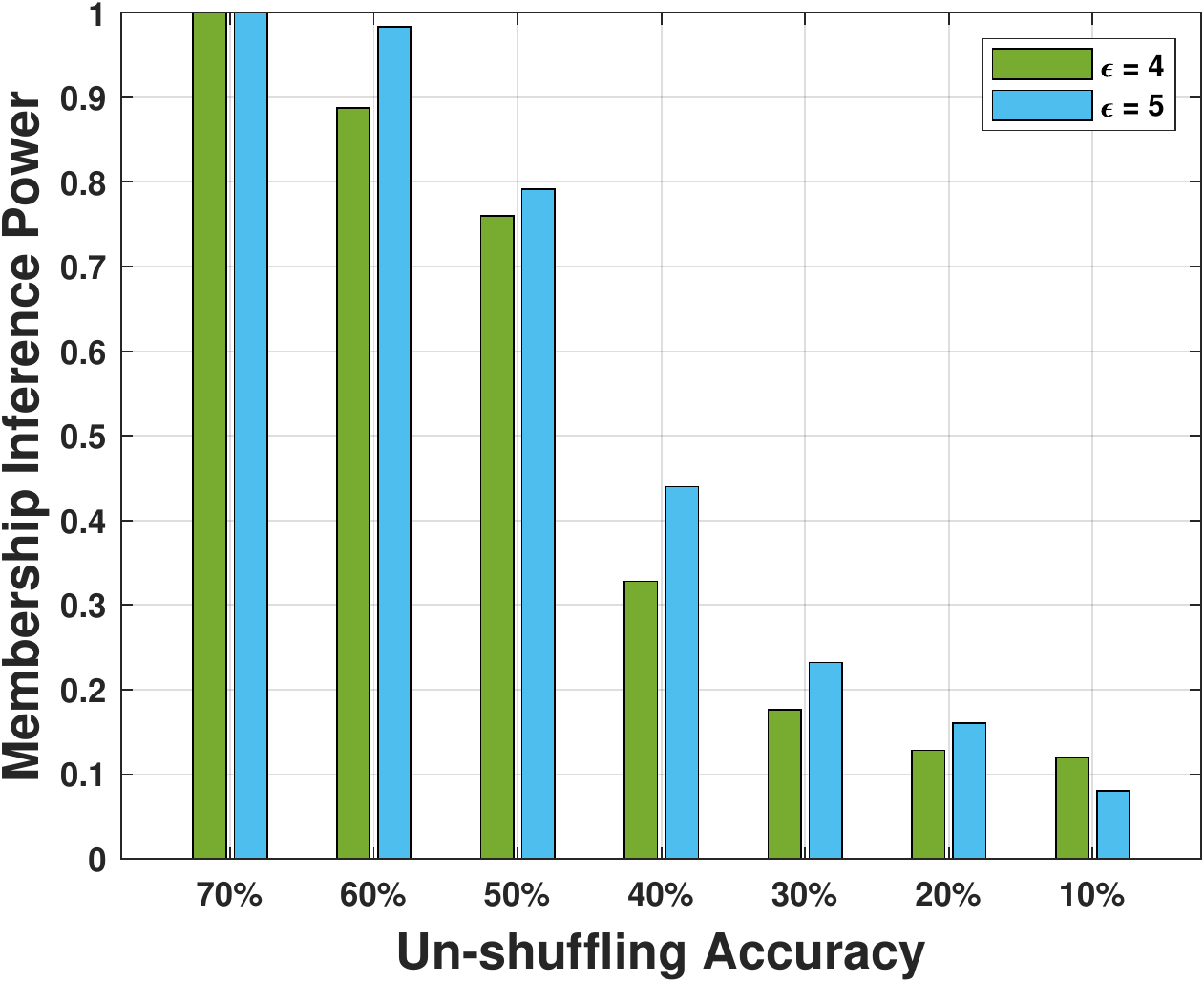}\vspace{-6pt}
\caption{\footnotesize{Power of membership inference attack based on hamming distance for different $\epsilon$ values considering various un-shuffling accuracy results.}}
\label{fig:Membership_inference}
\vspace{-10pt}
\end{wrapfigure}

Initially, we explored the effect of synthetic user addition (when different number of synthetic users are added to $M^i$) to the un-shuffling accuracy when the researchers randomly pick the shared SNPs (brown colored line in the figure). 
We observed a steady decrease in the un-shuffling accuracy as the number of added synthetic samples increases. The un-shuffling accuracy drops below 50\% when more than $400$ synthetic samples (which corresponds to $80\%$ of the original dataset) is added to $M^i$. 
Then, we let the researchers pick SNPs with close MAFs in $M^i$ (with an average distance of 0.00017, compared to 0.0329 for randomly picked SNPs) and as expected, we observe an additional decrease in the un-shuffling accuracy (green colored line in the figure).
This is because, when the MAF values of the SNPs in $M^i$ are close to each other, the probability of the server to incorrectly infer the ID of a SNP during un-shuffling (using the proposed greedy algorithm) increases.
Note that picking SNPs with close MAFs, generally provides slightly lower utility for the shared data.
We also observed that when the size of $I'$ (the set of shared SNP IDs that is known by the server) is larger than the size of $I$ (the set of shared SNP IDs in $M^i$), the un-shuffling accuracy further decreases, as expected (red and cyan colored lines in the figure). 
Next, we explored the effect of different values of the privacy parameter $\epsilon$ (i.e., when different amount of noise is added to the entries in $M^i$ to achieve the proposed LDP variant). We observed that addition of noise significantly decreases the un-shuffling accuracy (violet, orange, and blue colored lines in the figure).

Finally, we show the power of membership inference attack at the server for different $\epsilon$ and un-shuffling accuracy values in Figure~\ref{fig:Membership_inference}. 
We observed that when the server has an un-shuffling accuracy of at least 70\%, the power of membership inference is $1$ for different $\epsilon$ values. However, as we show in Figure~\ref{fig:Unshuffling_accuracy}, such a high un-shuffling accuracy is only possible when no synthetic samples or noise is added to the metadata. 
Note that the un-shuffling accuracy on the x-axis in Figure~\ref{fig:Membership_inference}, corresponds to a virtual horizontal line that can be drawn in Figure~\ref{fig:Unshuffling_accuracy}. Then, using any of the alternatives that lies below that line would then be sufficient to achieve a power of membership inference that is less than the corresponding one shown in the y-axis of Figure~\ref{fig:Membership_inference}. 
For instance, to have the power of membership inference less than $0.5$, one should use one of the techniques that provide an un-shuffling accuracy of 40\% or below in Figure~\ref{fig:Unshuffling_accuracy}. 
In this case, all techniques that include selection of SNPs with close MAFs in $M^i$ and adding at least $300$ synthetic samples (which corresponds to $60\%$ of the original dataset size) would be sufficient (i.e., all lines except for the brown colored one in Figure~\ref{fig:Unshuffling_accuracy}). 
As also shown in Table~\ref{table:kinship_accuracy}, almost all these techniques provide high kinship accuracy, and hence utility.  
We also observed that for all un-shuffling accuracy values, the proposed scheme provides higher privacy (i.e., smaller membership inference power) than the baseline risk, which occurs due to the sharing of summary statistics about the dataset (e.g., MAF values).  

Overall, given a fixed privacy goal (e.g., in terms of maximum membership inference risk), researchers need to optimize the accuracy of the kinship calculation and the computational load at the server (along with the bandwidth).
To do so, first, given the maximum membership inference risk, using the results of Figure~\ref{fig:Membership_inference}, researchers decide the maximum un-shuffling accuracy that can be tolerated. 
Next, using the results of  Figure~\ref{fig:Unshuffling_accuracy} and given the maximum un-shuffling accuracy to be tolerated, researchers identify the potential techniques that can be used in terms of number of synthetic samples to be added and the privacy parameter to achieve LDP. Finally, considering (i) the kinship accuracy due to each potential technique (from the result of Table~\ref{table:kinship_accuracy}) and (ii) computational load and bandwidth (depending on the number of synthetic samples), researchers decide on the technique and parameters to be used. 
\vspace{-5pt}
\section*{Conclusion}\label{sec:conclusion}
\vspace{-10pt}

In this work, we have proposed an efficient and effective privacy-preserving technique to identify correlated samples across distributed genomic datasets. 
Focusing on kinship relationships, we showed via experiments on real-life genomic data that the proposed scheme simultaneously provides high privacy guarantees for the research participants and high accuracy for the identification of kinship relationships between individuals across different research datasets. It also allows the researchers to customize system parameters to control the trade-off between the accuracy (for correct identification of the kinship relationships) and computational efficiency (i.e., computational load at the server). 
The proposed scheme will facilitate collaborative research results to be obtained using high quality data without violating privacy of the research participants. 
\vspace{-5pt}
\makeatletter
\renewcommand{\@biblabel}[1]{\hfill #1.}
\makeatother

\bibliographystyle{vancouver}
\bibliography{amia}  

\end{document}